\documentclass[letter]{aa} 

\DeclareHookRule{begindocument}{hyperref}{before}{nameref}
\usepackage{graphicx}
\usepackage{txfonts}

\usepackage{natbib}
\bibpunct{(}{)}{;}{a}{}{,} 

\usepackage[colorlinks=true,citecolor=blue,urlcolor=blue]{hyperref}

\makeatletter
\AtBeginDocument{%
  \@ifundefined{serieslogo}{}{}
  \@ifundefined{extraheadheight}{}{}
}
\makeatother

\let\linenumbers\relax

\nolinenumbers
\begin{document}

   \title{Polarization echoes from past nuclear activity in the quasi-periodic eruption source GSN~069}

   \author{B. Ag\'is-Gonz\'alez\inst{1,2,3}, D. Hutsem\'ekers\inst{3}, I. Liodakis\inst{1}, S. Cazzoli\inst{2}, D. Sluse\inst{3}, G. Miniutti\inst{4}, I. M\'arquez\inst{2}, J. Masegosa\inst{2}, F. Marin\inst{5}, J. A. Acosta-Pulido\inst{6}\fnmsep\inst{7} \and C. Ramos Almeida\inst{6}\fnmsep\inst{7}}
      
   \institute{
   Institute of Astrophysics, FORTH, N.Plastira 100, Vassilika Vouton, 70013 Heraklion, Greece
   \and
   Instituto de Astrof\'isica de Anadaluc\'ia, IAA-CSIC, Glorieta de la Astronom\'ia, s/n, 18008, Granada, Spain.
   \and
   Institut d'Astrophysique et de G\'eophysique, Universit\'e de Li\`ege, All\'ee du 6 Ao\^ut 19c, 4000 Li\'ege, Belgium
   \and 
  Centro de Astrobiología (CAB), CSIC-INTA, Camino Bajo del Castillo s/n, 28692 Villanueva de la Cañada, Madrid, Spain
    \and
     Universit\'e de Strasbourg, CNRS, Observatoire Astronomique de Strasbourg, UMR 7550, F-67000 Strasbourg, France
   \and 
   Instituto de Astrof\'isica de Canarias, Calle V\'ia L\'actea, s/n, E-38205, La Laguna, Tenerife, Spain
    \and
   Departamento de Astrof\'isica, Universidad de La Laguna, E-38206, La Laguna, Tenerife, Spain}

   \date{Received; accepted}

  \abstract
   {X-ray quasi-periodic eruptions (QPEs) are repeating, high-amplitude, soft X-ray bursts observed from the nuclei of a dozen nearby low-mass galaxies. Their origin remains a major puzzle in the physics of accretion variability. Observational data indicate that X-ray and/or optical tidal disruption events (TDEs) may precede QPE detections. Although both kinds of outburst are driven by supermassive black holes, they are more frequently detected in faded active galactic nuclei (AGNs), when the TDE is not happening in a dormant galaxy. In the case of the QPE discovery source, GSN~069, observations and simulations have revealed evidence of past nuclear activity, although it remains debated whether this activity arose from a past AGN phase or from an enhanced TDE rate.}
   {We investigated the origin of the past nuclear activity in GSN~069.} 
   {Past AGN activity imprints detectable polarization in optical light, due to the expected delay between direct and scattered light. On 6 September 2019, we targeted GSN~069 with VLT/FORS2 in both imaging polarimetry and spectropolarimetry modes so that its optical polarization could be investigated while the first detected QPE phase was still active.}
   {We  measured a rising polarization, from $\sim 0\%$ to $\sim1.5\%$, as moving away from the nucleus of GSN~069. This rise is probed to be intrinsic to the central engine, confirming the already detected extended emission line region (EELR) by integral field unit data.} 
   {The increasing radial polarization demonstrates a switched-off nucleus. The polarization angle traces an axis aligned with elongated [OIII], [NII], and H$\alpha$ gas distributions, revealing an EELR that may be consistent with relic polarization cones, therefore suggesting the presence of a torus-like structure in the past. Thus, optical polarization echoes geometrically favor a faded AGN as the origin of the EELR rather than a past elevated TDE rate, although the latter cannot be excluded.}


\titlerunning{Polarization echoes from past nuclear activity in the QPE source GSN~069}
\authorrunning{B. Ag\'is-Gonz\'alez et al.}
\maketitle
%
\section{Introduction}

Systematic multi-wavelength sky surveys have opened a new era in detecting nuclear transients (NTs), i.e., outbursts in galactic nuclei driven by accretion variability onto supermassive black holes. The recently discovered quasi-periodic eruptions (QPEs, \citealt{Miniutti2019Natur}) are striking, high-amplitude (several orders of magnitude), quasi-periodic X-ray bursts of variable duration (1--100 hours), recurring on timescales of hours to weeks. Their origin remains a mystery, though 12 sources have now been confirmed \citep{Miniutti2019Natur,Giustini2020, Arcodia2021Natur,Chakraborty2021,Quintin2023,Arcodia2024,Nicholl2024Natur,HernandezGarcia2025NatAs,Bykov2025,Chakraborty2025,Arcodia2025}. Some showed either X-ray (GSN069, \citealt{Miniutti2023}; XMMSL1 J024916.6-041244J0249 \citealt{Chakraborty2021}) or optical (AT2019qiz, \citealt{Nicholl2024Natur}; AT2019vcb, \citealt{Bykov2025}; AT2022upj, \citealt{Chakraborty2025}) tidal disruption events (TDEs) prior to the QPE detection. This TDE--QPE link has motivated the TDE+EMRI=QPEs model \citep{Linial2023,Linial2025}, where a stellar object, brought into the nucleus as an extreme mass-ratio inspiral (EMRI), passes twice per orbit through an existing accretion disk on a low-eccentricity orbit, reproducing the observed properties of QPE flares. In addition, TDE and QPE hosts also show a preference for extended emission line regions (EELRs) \citep{French2023,WeversFrench2024,Wevers2024}. 

GSN~069, like the other QPE sources, is X-ray unobscured \citep{Miniutti2013}, while its current X-ray emission is compatible with a compact TDE-like accretion disk \citep{Guolo2025} that switched on at least 8 years before the first QPE detection in December 2018. However, no optical broad emission lines (BELs) were detectable in deep long-slit \citep{Wevers2022} or in MUSE integral field unit data \citep{Wevers2024}. The MUSE data revealed a $\sim10$kpc EELR,  photoionized by a nonstellar continuum exceeding two to three times the current nuclear luminosity, pointing to nuclear activity that shut down  thousands of years ago \citep{Wevers2024}. This past energy budget could be accomplished by two different scenarios: either a faded AGN \citep{French2023, Wevers2024} or a past elevated TDE rate as photoionizing engine \citep{Mummery2025}. In both cases, considering that the recent TDE-like activity is unable to sustain mature BELs, only narrow-emission lines (NELs) and the EELR remain observable as fossil records of the past nuclear activity. \cite{Hutsemekers2019} have proposed that a past bright AGN phase imprints an echo in polar-scattered light, thus producing detectable polarization. For this letter we tested this scenario for GSN069 by resolving the distribution of the polarized flux until $\sim2$kpc from the nucleus with VLT/FORS2 photo- and spectropolarimetry data taken on 6 September 2019, when the first detected QPE phase of GSN069 was still active, after the decay of a TDE-like X-ray flare first detected in 2010 and preceding a second X-ray rebrightening possibly associated with a repeated TDE. \citep{Miniutti2023}.

\section{Direct spectrum and polarization of GSN069}
\begin{figure}
\centering
   \includegraphics[trim={0mm 3mm 0mm 5mm}, clip, width=\linewidth
   ]{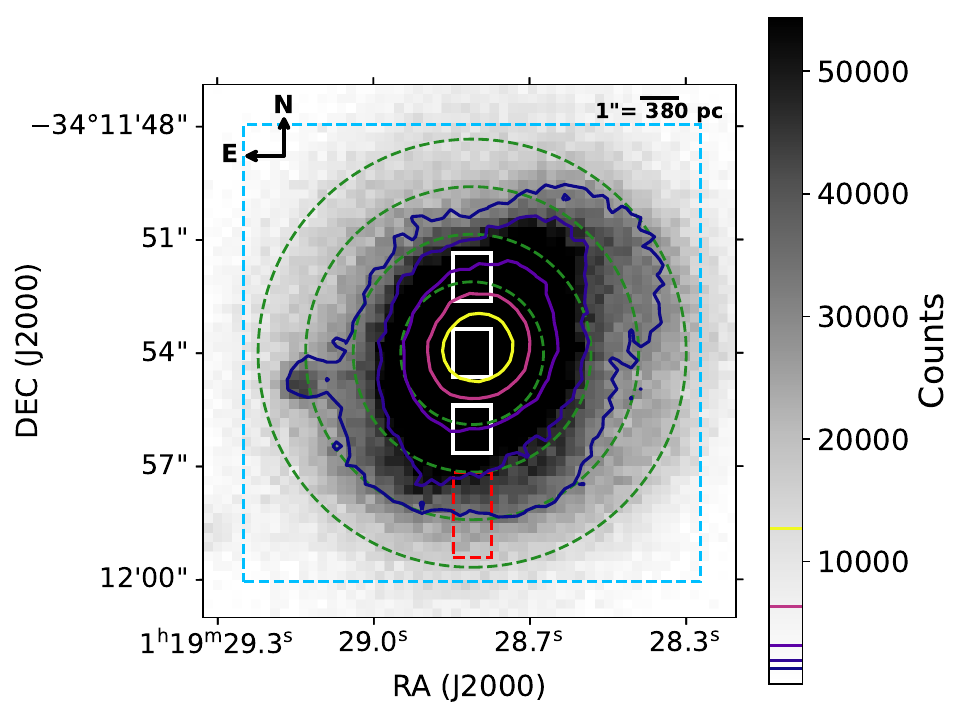}
   \caption{Acquisition image of GSN 069. Flux contours follow the shown color scale. The green dashed circles mark IPOL aperture areas with radii of $\sim3\sigma$ (2.1"), $\sim5\sigma$ (3.5"), $\sim7\sigma$ (4.9"), and $\sim9\sigma$ (6.3"), where $\sigma$ is the seeing value during IPOL observations, which remained stable at $\sim$ 0.7\arcsec. The white boxes indicate the north, nucleus, and south regions used for PMOS extractions (see text), while the red box show the southern spot extraction region. The blue dashed box outlines the same $12''\times12''$ area as the Chandra image of GSN~069 shown in panel~d of Extended Data Fig. 1 in \cite{Miniutti2019Natur}. The spot is not visible in our optical FORS2 IPOL images, but appears in those Chandra observations and in the MUSE continuum-subtracted H$\alpha$ image (Fig. \ref{fig:muse} and Appendix \ref{append:Spot}).}
    \label{fig:integration_areas_contours}%
\end{figure}

Details of the observations and data analysis are provided in Appendix~\ref{appendix:reduction}. We  found a bright feature located $4.25\arcsec$ south of the nucleus (hereafter the southern spot) in our FORS2 PMOS spectral images (Fig.~\ref{fig:spot}). The southern spot is also present in X-ray Chandra data (Extended Data Fig. 1 in \cite{Miniutti2019Natur}) and in the MUSE continuum-subtracted H$\alpha$ image (Fig.\ref{fig:muse}), although it is not resolved in our FORS2 IPOL optical images (Fig.~\ref{fig:integration_areas_contours}). Therefore, in addition to the direct spectrum and polarization of the nucleus, we also analyzed the spectra extracted from the southern spot and from two regions in the host galaxy, $\sim1.75\arcsec$ north and south of the nucleus (see Fig.~\ref{fig:integration_areas_contours} and Appendices~\ref{appendix_subsec:direct_spectra} and~\ref{append:host}). At the distance of GSN~069 ($z=0.0181$; \citealt{Miniutti2013}), 1\arcsec $\simeq$ 380~pc.

\subsection{Direct spectrum} 
\label{subsec:direct_spectrum}

Figure~\ref{fig:direct_spectra} shows the optical direct spectra of the GSN~069 nucleus and the southern spot, while Fig.~\ref{fig:HOST_spectra} presents the north and south extractions. After subtracting the computed starlight model, no continuum emission remains or broad Balmer line components arise in any extraction (see Appendices~\ref{appendix_subsec:direct_spectra}, \ref{append:broad_lines}). Nonetheless, the standard emission-line ratios used in BPT diagnostics \citep[e.g.,][]{Kewley2001} classify GSN~069 as an AGN at the nucleus, southern spot, and both host-galaxy regions. Combined with the absence of X-ray obscuration, this indicates that the current nuclear activity has shut down and it is unable to sustain a persistent broad-line region. We are therefore observing a relic narrow-line region compatible with either a faded AGN phase \citep{Wevers2024} or a past enhanced rate of TDEs \citep{Mummery2025}.

\subsection{Imaging polarimetry} 
\label{subsec:imagingpol}

\begin{figure}[]
   \centering
   \includegraphics[trim={0mm 3.4mm 0mm 2.9mm}, clip, width=0.8\linewidth]{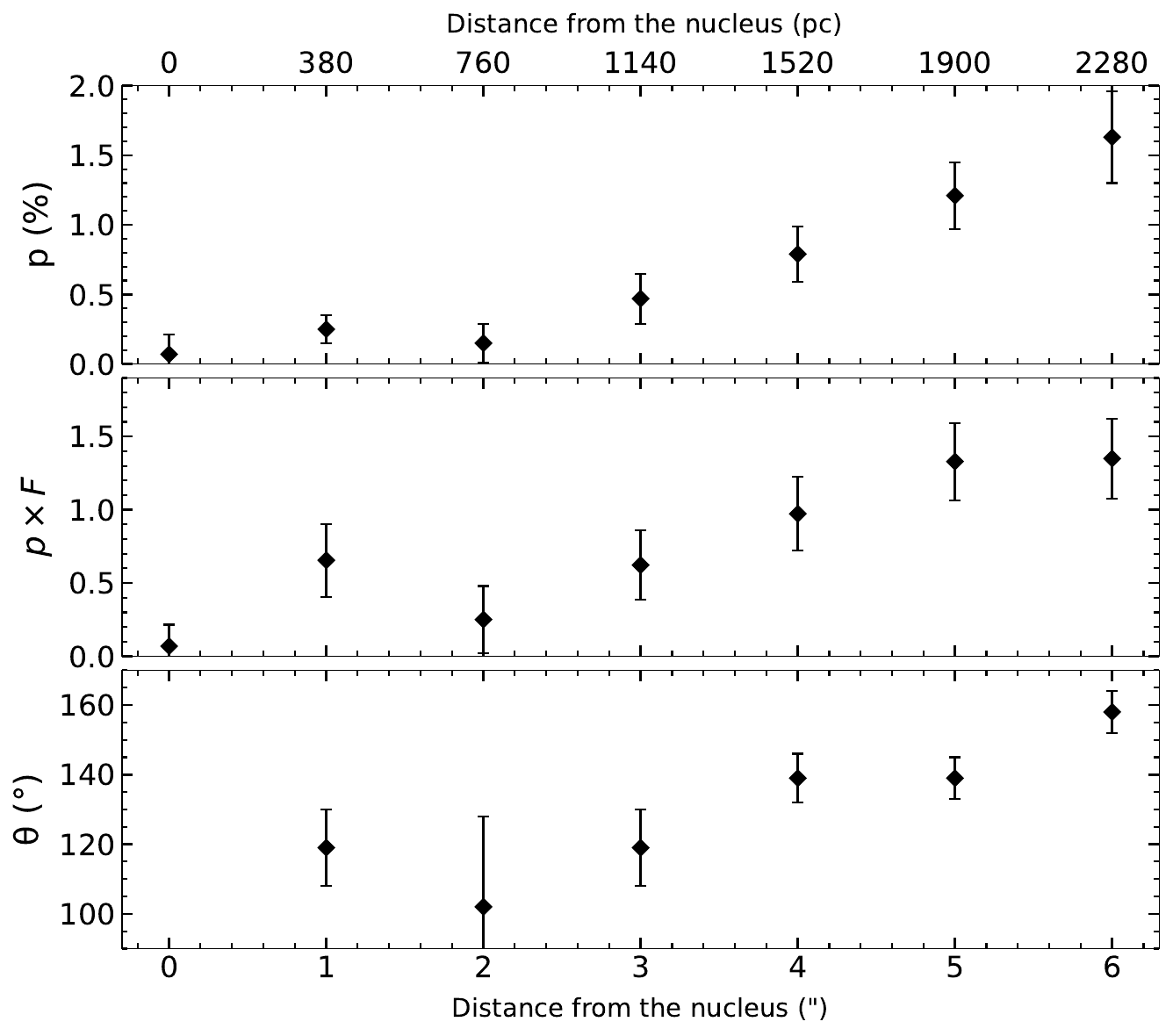}
   \caption{Polarization degree (p), polarized flux ($p\times F$in arbitrary units), and polarization angle ($\theta$) measured in the V band, through an aperture of 4 pixels of  diameter (1\arcsec) centered on the nucleus, and in 4 pixel wide annuli of increasing radius (1\arcsec $\simeq$ 380~pc). The polarization degree is not corrected for the bias. The first bin is circular (nucleus) showing $p$ compatible with zero and an undefined $\theta$.}
   \label{fig:asymmetry1}%
\end{figure}

With our IPOL data, we found a good compromise between capturing the full nuclear emission and minimizing host-galaxy contamination by choosing an aperture radius of approximately three times the value of the seeing, i.e., $\sim 3\sigma$;   $\sigma$ remained stable at around 0.7\arcsec during the observations. This yielded a polarization degree of $p=0.15\pm0.11\%$ and a polarization position angle of $\theta=126^\circ\pm20^\circ$. Following \cite{Plaszcsynski2014}, the debiased polarization degree is $p_{0}=0.12\pm0.11\%$, certainly compatible with null polarization. However, as we increased the aperture radius, $p$ also increased and $\theta$ became better defined (see Table \ref{tab:IPOL}). Figure~\ref{fig:integration_areas_contours} shows the different integration circular apertures with radii $3\sigma$ (2.1\arcsec), $5\sigma$ (3.5\arcsec), $7\sigma$ (4.9\arcsec), and $9\sigma$ (6.3\arcsec) on the acquisition image of GSN~069. The radial trend is evident in Fig.~\ref{fig:asymmetry1}, where $p$ is measured in annular apertures (represented in Fig.\ref{fig:muse}) instead of circular apertures to exclude the nucleus, revealing a net rotation of $\theta$ with distance. The radio emission of GSN~069, consistent with a compact outflow associated with the ongoing X-ray TDE activity \citep{Goodwin2025}, prevents an estimation of its symmetry axis, and thus a comparison with $\theta$ is not feasible. Figure 1 in \cite{Guolo2025}, which decomposes the radial profile using HST F606W images (similar wavelength coverage to our data; see Appendix \ref{appendix_subsec:impol_obs_red}), suggests that our smallest aperture, 2.1\arcsec, isolates the switched-off nucleus, while larger apertures encompass the EELR/ionization cones. This scenario is further analyzed in the following sections.

\begin{table}
\caption{Imaging polarimetry results for different aperture radii, $r_{aper}$, in seeing units, where $\sigma=0.7\arcsec$. $p_{0}$ is the debiased polarization degree.}
\label{tab:IPOL}      
\centering            
\begin{tabular}{c c c c c c}  
\hline\hline       
$r_{aper}$ &   $q$    &   $u$    &   $p$  & $p_0$ & $\theta$ \\    
($\sigma$) & (\%)  & (\%)   & (\%) & (\%) & ($\degr$)   \\    
\hline                        
3           & -0.05 & -0.15 & $0.15\pm0.11$ & 0.12 & $126\pm20$ \\     
5           & -0.08 & -0.34 & $0.35\pm0.10$ & 0.34 & $129\pm8$ \\
7           & -0.00 & -0.54 & $0.54\pm0.10$ & 0.53 & $135\pm5$ \\
9           &  0.11 & -0.76 & $0.77\pm0.10$ & 0.76 & $139\pm4$\\
\hline
\end{tabular}
\end{table}

Unpolarized starlight from the host galaxy could dilute the intrinsic nuclear polarization, as it is coupled to the nuclear flux emission (e.g., \citealt{MillerAntonucci1983, Marin2018}). We  tried to estimate the host contribution at each aperture radius by using our IPOL acquisition image (60 s integration, no stripped mask used). We modeled GSN~069 as the sum of (1) a 2D Gaussian with the FWHM fixed to the seeing value, representing the nucleus as a point source, and (2) a convolved Sersic profile \citep{Sersic1968} with n=4, i.e., deVaucouleurs law \citep{deVaucouleurs1948} representing the host galaxy. This general model suggests that the host intensity exceeds the nuclear contribution by a factor of $\sim10$ at our smaller aperture radius, i.e., 2.1\arcsec, consistent with the radial decomposition by \cite{Guolo2025}. The low spatial resolution of our acquisition image, together with the coupling of the nuclear intrinsic emission with the host contribution, prevents meaningful constraints on a proper host correction. Nevertheless, the observed increasing polarization with radius implies that, despite gathering more and more host galaxy inside the aperture, there is no  depolarizing effect, even at the location of the star-forming ring revealed by MUSE data \citep{Wevers2024} at radius $\sim2$\arcsec.

\subsection{Spectropolarimetry}
\label{subsec:specpol}

\begin{figure}
\centering
\includegraphics[trim={0mm 3mm 0mm 0mm}, clip, scale=0.32]{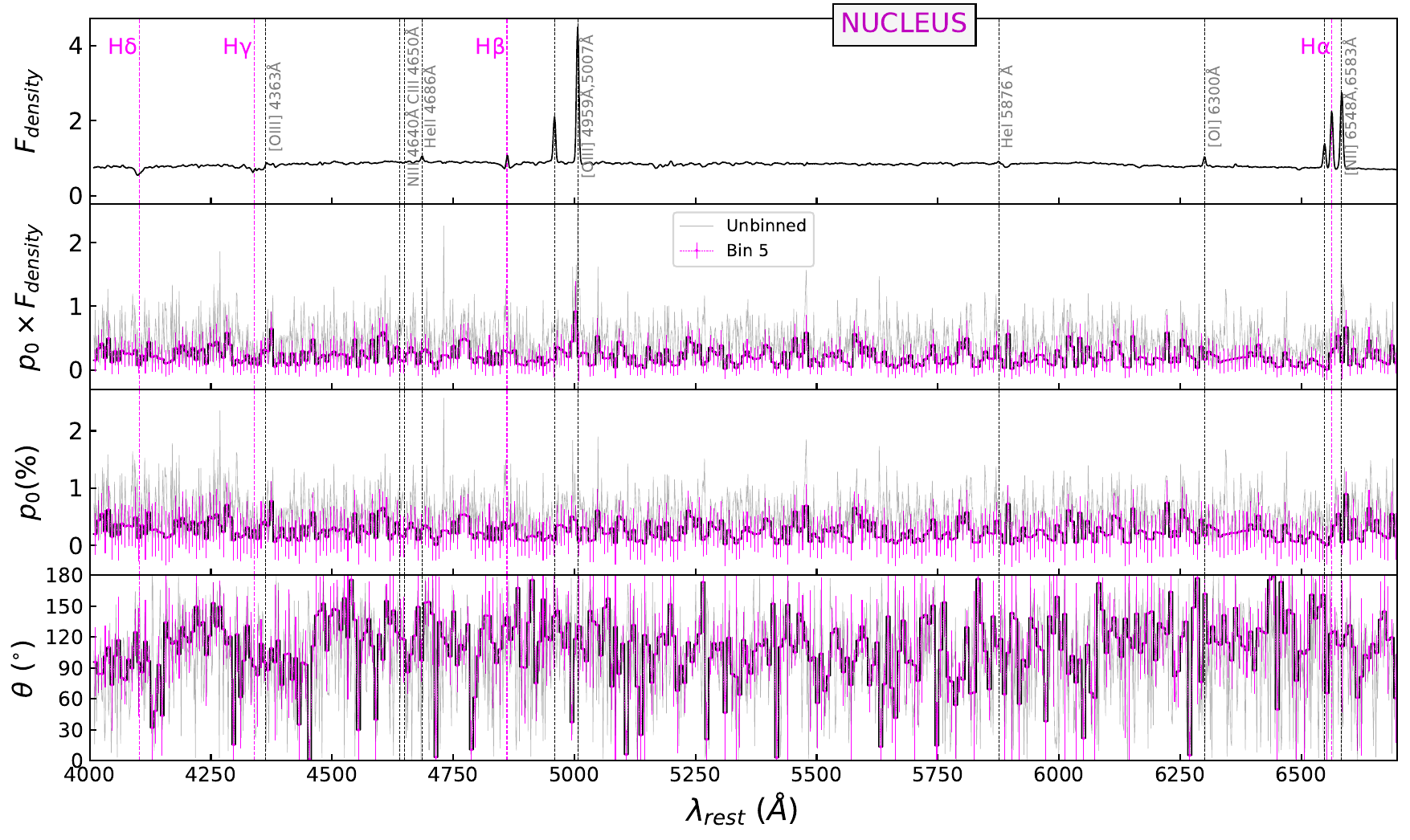}
\includegraphics[trim={0mm 3mm 0mm 0mm}, clip, scale=0.32
]{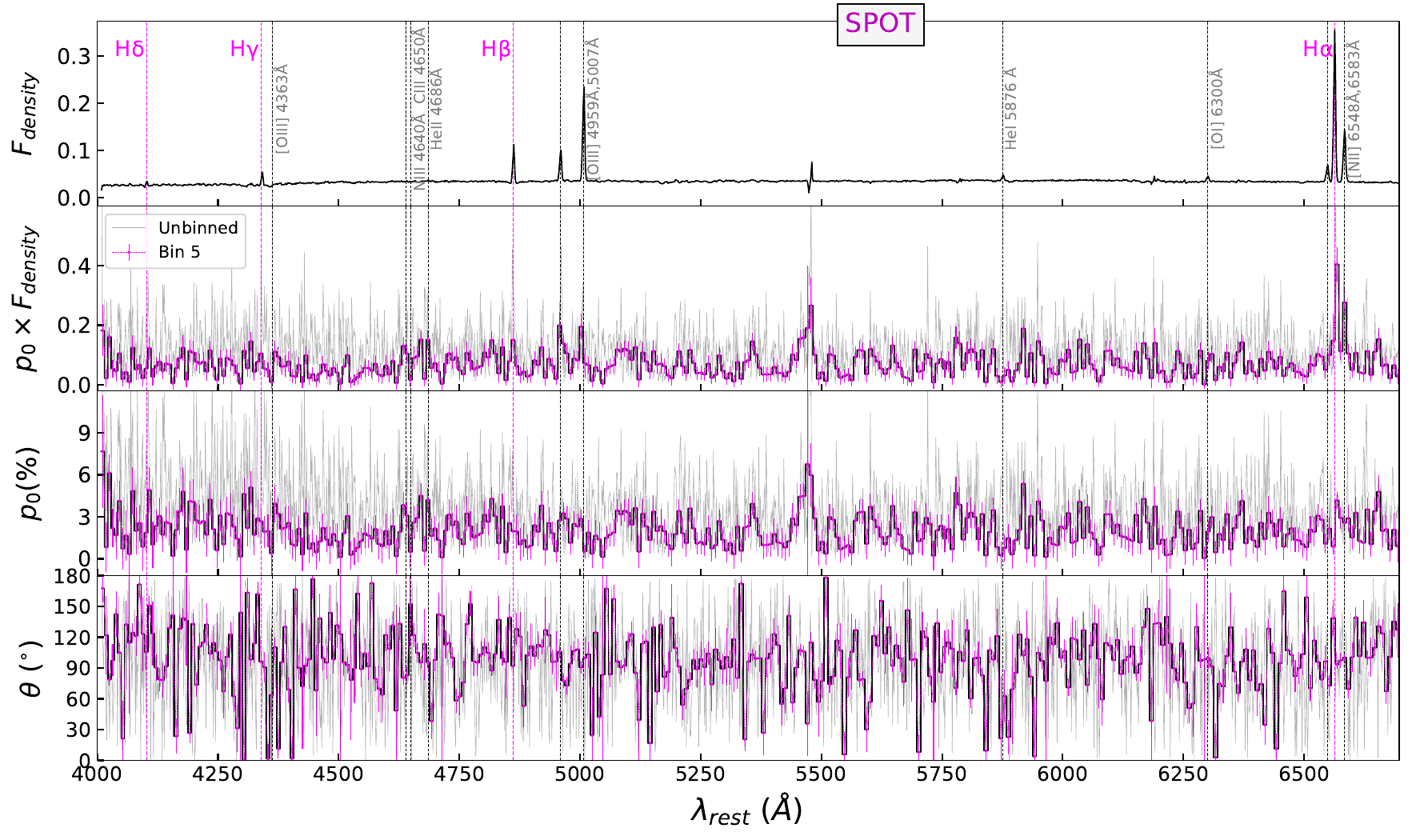}
   \caption{Flux density in arbitrary units ($F_{density}$), debiased polarization degree ($p_{0}$), polarized flux ($F_{density}\times p_{0}$), and polarization angle ($\theta$) for the nucleus (upper panel) and southern spot (lower panel).}
   \label{fig:specpolGSN069}%
\end{figure}

Figure~\ref{fig:specpolGSN069} shows the direct spectrum, the polarized spectrum, the debiased polarization degree, and the polarization position angle for the nucleus and the southern spot. Figure \ref{fig:specpolGSN069host} shows the same data for the two regions north and south of the nucleus. In general, and particularly in the nucleus, the polarization signal is low, yielding a low signal-to-noise ratio (S/N). The noise hampers accurate line identification in the polarized spectra, and $\theta(\lambda)$ remains poorly define due to the low polarization degree.  

\begin{figure}
   \centering
   \includegraphics[trim={0mm 3.5mm 0mm 2.9mm}, clip, width=0.88\linewidth]{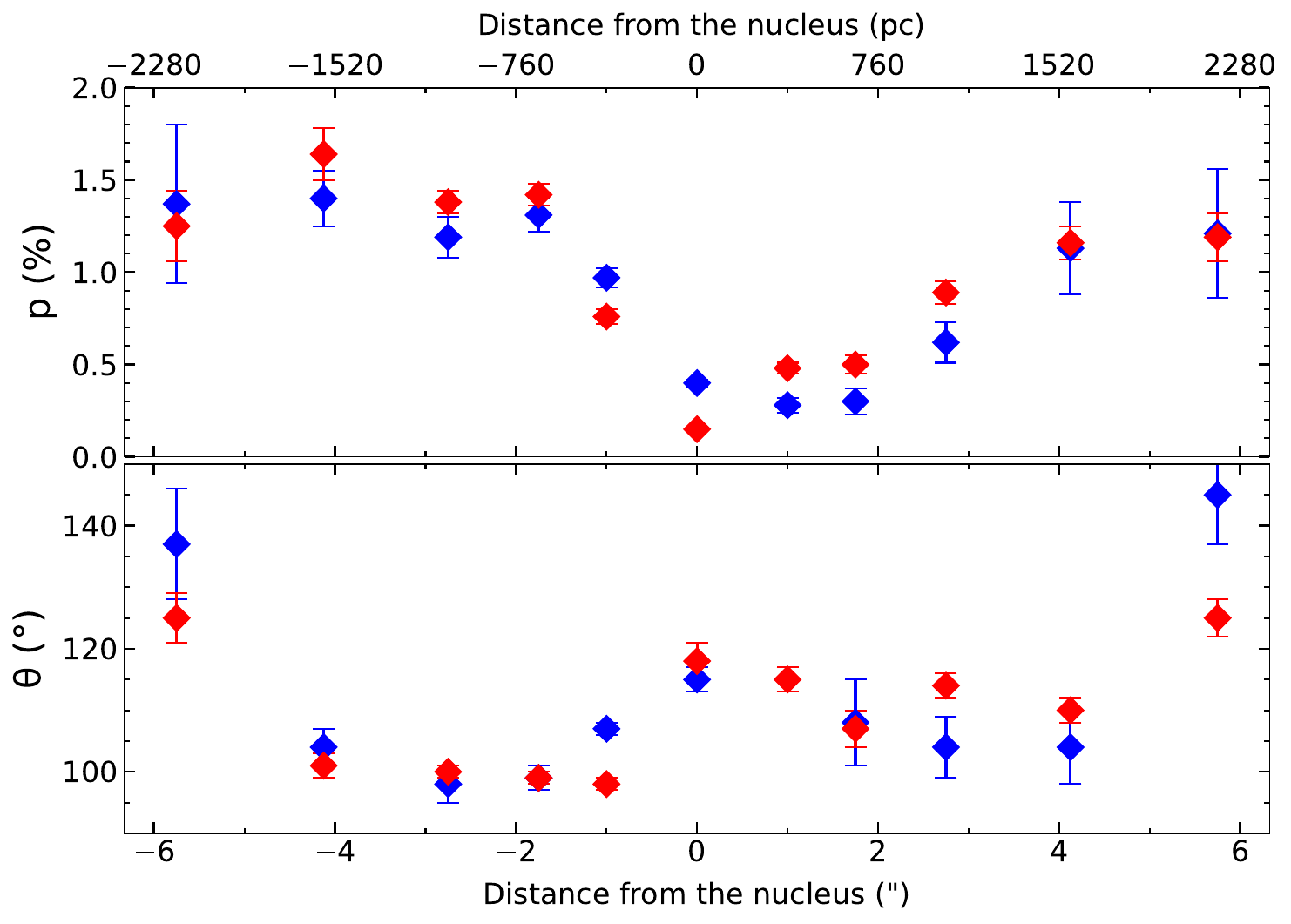}
   \caption{Polarization degree and angle measured in spatial bins (subslits) along the slit. The polarization degree is not corrected for the bias. The bin centered on the nucleus is 5 spatial pixels (1.25\arcsec) long. The subsequent bins are 3, 3, 5, 6, and 7 pixels long. Positive distances are toward the north, negative ones toward the south (1\arcsec $\simeq$ 380~pc). Measurements integrated over the continuum wavelength ranges 3800-5000 \AA < and 5400-6600 \AA\ are shown in blue and red, respectively. The bin at $-$4.1\arcsec contains the southern spot.}
   \label{fig:asymmetry2}%
\end{figure}

We averaged the Stokes parameters $q$ and $u$ between 5100 and 6500 $\AA$ (line-free range) to compute the corresponding continuum polarization degree $p_{5100-6500}$ (not debiased to compare with Fig.\ref{fig:asymmetry1} and Fig.\ref{fig:asymmetry2})  and polarization angle $\theta_{5100-6500}$, assigning $\sigma_p$ as the 1$\sigma$ error and adopting $\sigma_\theta=28.65\degr\frac{\sigma_p}{p}$ \citep{Bagnulo2009}. We obtained respectively $p_{5100-6500}=0.22\%\pm0.02\%$ and $\theta_{5100-6500}=111^\circ\pm2^\circ$ for the nucleus, $0.35\%\pm0.04\%$ and $105.00^\circ\pm3^\circ$ for the north, $1.20\%\pm0.05\%$ and $102^\circ\pm1^\circ$ for the south, and $1.5\%\pm0.11\%$  and $98^\circ\pm2^\circ$ for the southern spot. In agreement with imaging polarimetry, polarization increases as we move away from the nucleus. Figure\ref{fig:asymmetry2} shows the polarization measured in spatial bins (see Fig.\ref{fig:muse} for their exact location). That binning reveals an asymmetry: $p$ is higher in the south than in the north, with the southern spot (contained in the $-4.1\arcsec$ bin) matching the south region within the errors. $\theta$ rotates between the north and south and the southern spot regions. The bins at $\pm6\arcsec$ deviate from this trend with $\theta \sim 140^\circ$ in both north and south. As seen in Fig.~\ref{fig:muse}, these bins lie near the edge of the EELR, where a change in the dominant polarization mechanism may occur.

As in the imaging polarimetry analysis (Sect.~\ref{subsec:imagingpol}), we attempted to estimate the contribution of the host galaxy to the spectropolarimetric measurements. We modeled the stellar continuum and absorption features, as done for the direct spectra, for each ordinary and extraordinary beam. Since the nuclear continuum is almost completely removed after starlight subtraction, residual noise or slight mismatches in the scaling of the stellar model introduce spurious flux values, yielding artificially high $p(\lambda)$ and meaningless $\theta(\lambda)$. Fortunately, the polarized flux, $p \times F$, should remain unaffected by starlight dilution \citep{Tran1995_II}, provided the starlight is entirely unpolarized.

We also searched for broad Balmer emission lines in the polarized flux via  line-profile fitting (see Appendix \ref{append:broad_lines}). Although the S/N is not optimal, no broad components are detected in the nucleus, the spot, or the two other regions. In contrast, the brightest narrow emission lines ([OIII], H$\alpha$+[NII]) are seen in the polarized flux, especially in the southern regions.

\section{Discussion}

Both photo- and spectropolarimetry show that the polarization of GSN~069 is null or very small at the nucleus, but gradually increases up to 1.5\% at distances of about 2 kpc from the galaxy center. The decrease in polarized flux toward the nucleus (Fig.~\ref{fig:asymmetry1}) support a slow dimming of the source, where the time delay expected between direct and scattered light produces an echo of past nuclear activity \citep{Hutsemekers2019}, either by a faded AGN or by a past elevated rate of TDEs. Non-null polarization degrees (p) in annuli centered on the nucleus indicate that the polarization is not perfectly centrosymmetric, i.e., the scattering regions are not isotropically distributed and/or homogeneous, consistent with the non-axisymmetric morphology of the EELR revealed by the MUSE data of GSN~069 \citep{Wevers2024}. The polarization angle ($\theta$) measured in annuli at $\sim$4-5\arcsec ($\sim$1.5-2 kpc) from the nucleus (Fig.\ref{fig:asymmetry1}) and within a circular aperture of $\sim $6.3\arcsec($\sim2$ kpc) (Table \ref{tab:IPOL}) reveal a dominant polarization direction of $\sim140^\circ$. Assuming this scattering lead by the EELR, then the EELR axis should be perpendicular to $\theta\sim140^\circ$, where $\theta$ is measured with respect to the north--south (N-S) direction (the IPOL adapter position). To further investigate the EELR morphology, we computed H$\alpha$, [OIII]$\lambda5007$, and [NII]$\lambda6583$ line maps from archival GSN~069 MUSE observations (see Appendix \ref{append:Spot}). As shown in Fig.~\ref{fig:muse}, the EELR axis, as inferred from $\theta$, aligns with elongated gas distributions seen in the NE area of the three emission-line maps. This suggests a preferred direction in the illuminating photon field. This geometry is more naturally explained by fossil polarization--ionization cones shaped by a faded AGN and its long-lived torus structure, rather than by past episodes of enhanced TDE activity, which are not expected to maintain any sustained collimating structure (an arbitrary aperture for the polarizaton--ionization cones of 60\degr is drawn in Fig. \ref{fig:muse} to guide the eye). This torus-shaped, biconical pattern favor the faded AGN as the origin for the EELR. However, because this geometry remains tentative, our results cannot rule out the past elevated TDE rate scenario.

The spectropolarimetry in the southern regions gives $\theta\sim100^\circ$ (Fig.~\ref{fig:asymmetry2}), i.e., roughly perpendicular to the hypothetical line connecting the nucleus with the scattering regions within our slit. Since scattering generates a polarization angle perpendicular to the photon's last direction of flight before being scattered, the source that illuminated the southern scattering regions in the past should be the nucleus itself. The 10\degr departure to 90\degr in the south extractions together with a higher $\theta$ in the northern region can be explained again by the non-axisymmetric EELR reported by \cite{Wevers2024}.

\section{Conclusions}

\begin{enumerate}
      
      \item The nucleus of GSN~069 is compatible with null intrinsic polarization, but the polarization degree increases as we move away from the galaxy center (Fig.~\ref{fig:asymmetry1} and~\ref{fig:asymmetry2}), indicating polarization by scattering. In the southern regions, the polarization angle suggests  illumination by the switched-off nucleus, i.e., a polarization echo from past nuclear activity. In the northern region, the different polarization angle reveals nonhomogeneous or non-isotropic scattering regions. These results are   consistent with the detection of an EELR of non-axisymmetric morphology revealed by the MUSE data \cite{Wevers2024}.

      \item The proved intrinsic and radial increasing polarization demonstrates a switched-off nucleus compatible with both a faded AGN and a past elevated TDE rate. The polarization angle defines an axis that aligns with the gas distributions, revealing an EELR consistent with tentative relic polarization cones. This would suggest the presence of a torus in the past. Therefore, the polarimetry supports geometrically the faded AGN scenario, although TDEs cannot be ruled out.

      \item Broad-emission lines (BELs) are not detected in polarized light. Instead, narrow-emission lines (NELs) are seen in the polarized flux, more intense in the southern regions. 

      \item We found a bright spot at $\sim4.25\arcsec$ south of the nucleus. As all the analyzed direct spectra (see Fig.\ref{fig:integration_areas_contours}), it is classified as an AGN according to BPT diagrams (Fig. \ref{fig:direct_spectra}) and lacks continuum emission and BELs.
          
\end{enumerate}

\bibliographystyle{aa} 
\bibliography{gsn069} 

\begin{appendix} 

\section{Observations and data reduction}
\label{appendix:reduction}

The observations of GSN~069 were carried out on 6 September 2019 in visitor mode (Program ID: 0103.B-0791) using the European Southern Observatory (ESO) Very Large Telescope (VLT) equipped with the FOcal Reducer and low dispersion Spectrograph 2 (FORS2) mounted at the Cassegrain focus of Unit Telescope \#1 (Antu). Linear polarimetry was performed with the standard setup and using the  stripped mask. GSN~069 was positioned at the center of the field of view to avoid the significant off-axis instrumental polarization generated by the FORS2 optics \citep{PatatRomaniello2006,GonzalezGaitan2020}. All observations were carried out at an airmass $\lesssim 1.5$.  After bias subtraction and flat-field correction, every frame was processed to remove cosmic rays using the lacosmic Python package \citep{vanDokkum2012}.

\subsection{IPOL mode:  Imaging polarimetry}
\label{appendix_subsec:impol_obs_red}

Polarized images were secured with the filter V\_HIGH+114 ($\lambda_{c}=5550\ \AA$; FWHM$=1232\ \AA$), using the standard resolution collimator and $2\times2$ binning, which corresponds to a scale plate of  0.25" pixel$^{-1}$ and to a field of view (FoV) of $6.8'\times6.8'$ imaged onto two identical CCD detectors (CHIP1 and CHIP2). A sequence of four exposures of 30 s each ($4\times30$ s) was performed for GSN069 with the HWP rotated at $0\degr$, $22.5\degr$, $45\degr$, and $67.5\degr$ angles. In order to measure the polarization of bright field stars around GSN~069 we switched to a  $1\times1$ binning and $4\times3$~s exposures. The seeing remained stable and around 0.7" during the imaging polarimetry observations.

Data reduction of our imaging polarimetry data was performed using the procedures described in \cite{Sluse2005}. Polarized (BD-12~5133 and Vela1~95) and unpolarized (WD~1620-391) standard stars \citep{Cikota2017} were observed and reduced in the same way to check the whole reduction process. We measured a polarization degree for the unpolarized standard star of $0.06\pm0.06\%$, indicating that the residual instrumental polarization is negligible at the center of the field.

GSN~069 is situated at high galactic latitude ($b=-80.8\degr$), well away from the galactic plane. Thus, the expected spurious polarization induced by the interstellar medium (ISM) of our Galaxy should be negligible. Actually, following the galactic dust reddening and extinction tool provided by IRSA \footnote{https://irsa.ipac.caltech.edu/applications/DUST/} and \cite{Serkowski1975}, we found that $p_{ISM\ max}\lesssim 0.24\%$.

We have also estimated the polarization of the stars lying on the FoV of FORS2, assuming that $p_{ISM}$ should be constant along the $6.8'\times6.8'$ area covered by the instrument. There exists a radial instrumental polarization pattern across the CCD increasing to $\approx1.4\%$ at the edges, depending on the filter, caused by the optics of the dual-beam polarimeter of FORS2, and very well characterized \citep{PatatRomaniello2006,GonzalezGaitan2020}. Unfortunately, it was only possible to measure the polarization of a single field star. For that star, we found $p=0.38\pm0.06\%$ and $\theta=102^\circ\pm4^\circ$, which become $p=0.32\pm0.06\%$ and $\theta=105^\circ\pm4^\circ$ after applying the instrumental correction maps developed by \cite{GonzalezGaitan2020}, in reasonable agreement with the upper limit $p_{ISM\ max}\lesssim 0.24\%$. In conclusion, $p_{ISM}$ cannot be inferred, only upper limits. Therefore and hereafter (also in spectropolarimetry), we do not apply any correction to compensate for the galactic interstellar contamination. 

The polarization of GSN~069 was measured in various apertures to estimate the impact of the host galaxy. Figures \ref{fig:integration_areas_contours} and \ref{fig:muse} show the different regions/subslits defined along the manuscript. Table~\ref{tab:IPOL} gives the polarization quantities measured within the apertures drawn in Fig.~\ref{fig:integration_areas_contours}. The polarization degree has been debiased following the prescriptions of \cite{Plaszcsynski2014} and it is indicated as $p_{0}$ in Table~\ref{tab:IPOL} .

Finally, we tried to build polarization maps from the IPOL images, but the polarization signal along with the spatial resolution are not sufficient to compute reliable maps, which would require subpixel precision.

\subsection{PMOS mode:  Spectropolarimetry}

Polarized spectra were secured with the grism 600B+22 ($\lambda_{c}=4627\AA$; $\lambda_{range}=3300-6210\AA$) using a slit width of 1", which provides a dispersion of 0.75 $\AA$ pixel$^{-1}$ and a spectral resolution of $R=\lambda / \Delta\lambda=780$. To obtain simultaneous observations of H$\beta$ and H$\alpha$, we shifted the position of the slit on the EV2 detector to cover the spectral range $\sim3800-6700\AA$. The binning was $2\times2$ and the scale plate 0.25"/pixel. The seeing oscillated between 0.59" and 1.34" during spectropolarimetry, i.e., reaching values comparable to and higher than the slit width. 

A sequence of four exposures of 1800 s each, with the HWP rotated at $0\degr$, $22.5\degr$, $45\degr$, and $67.5\degr$ angles, constituted one observing block (OB). It was repeated three times, i.e., three OBs of about 2h15min each, so that the better signal-to-noise (S/N) could be achieved by (median) combining the different spectra and checking the consistency among the three sets. The instrument position angle was set to $PA=0\degr$ (that is, along the N-S direction).

In total, we obtained 24 spectra (from two orthogonal polarizations, four HWP angles, three OBs). The normalized Stokes parameters $q(\lambda)$ and $u(\lambda)$, corrected for the HWP chromaticity, as well as the linear polarization degree $p(\lambda)$ and the polarization position angle $\theta(\lambda)$ were computed from the individual, sky-subtracted, spectra following standard formulae \citep[e.g.,][]{Bagnulo2009,Marin2025}. 

The polarized standard star BD-12 5133 \citep{Cikota2017} was observed and reduced in the same way to check the whole reduction process in spectropolarimetry. The spectrophotometric standard star Feige 110 \citep{Moehler2014} was observed, also in spectropolarimetric mode, to generate a response curve and flux-calibrate the direct spectrum of GSN~069. 

Due to the good S/N, we extracted a central narrow region within a subslit of 5-pixel width (Fig.~\ref{fig:integration_areas_contours}), corresponding to 1.25" or $\sim1\sigma$, being $\sigma$ the average seeing value during spectropolarimetry. This represents the nucleus of GSN~069 with a minimal contribution of the host galaxy, having checked that wider/narrower regions/subslits do not provide any extra information. 

Even if the S/N in terms of direct flux is high, when the polarized signal is low, its corresponding probability distribution differs from Gaussian. To avoid the positive bias associated with the polarization degree $p$, a common practice is to rotate the plane $q-u$  by an angle $\theta$, being $\theta$ the inferred polarization angle, so that all the signal is contained in $q'$ and Gaussian errors can be considered. For our data, this rotation does not provide us with an improved view of our results, and we decided to show the standard polarization degree $p$ and position polarization angle $\theta$.

\subsection{PMOS mode:  Direct spectrum}
\label{appendix_subsec:direct_spectra}

\begin{figure}
   \includegraphics[trim={0mm 2.7mm 0mm 2.9mm}, clip, width=0.5\textwidth]{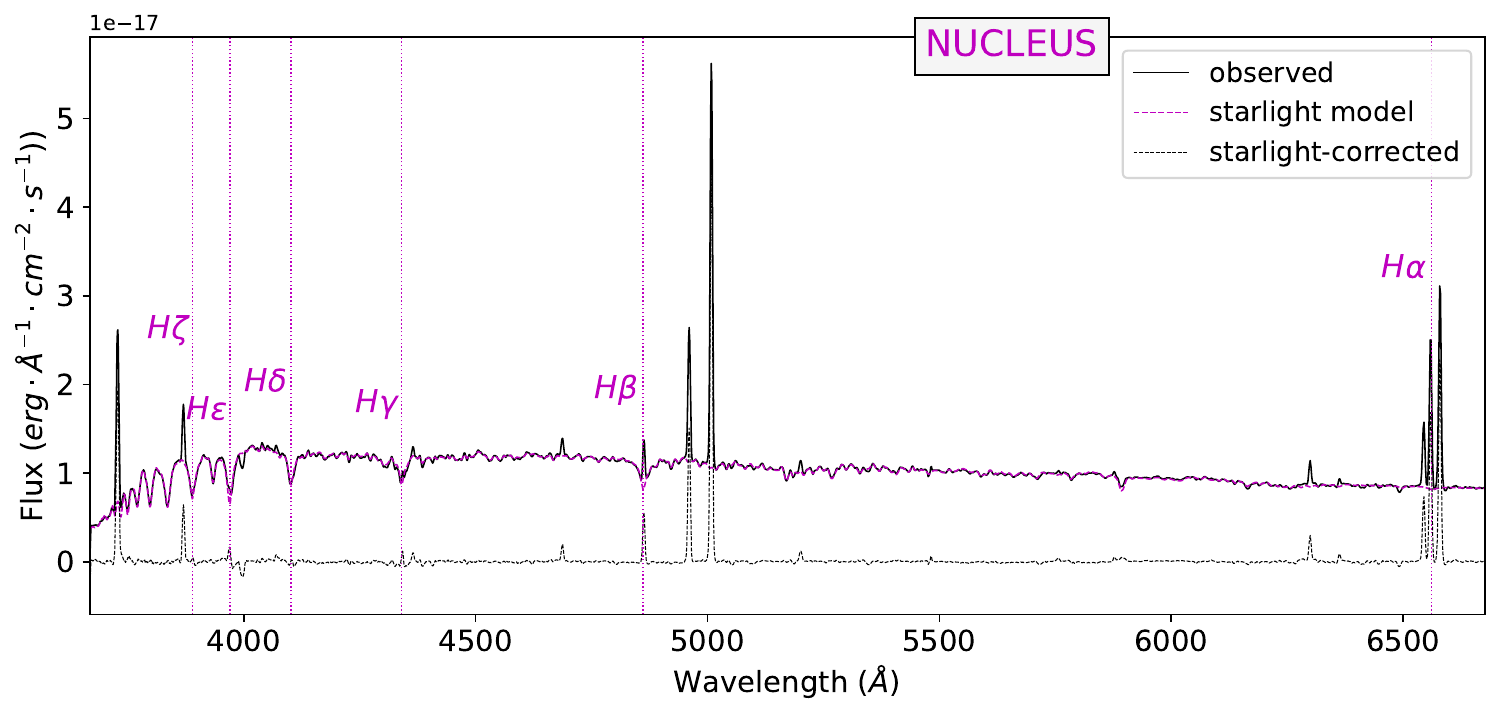} 
   \includegraphics[trim={0mm 2.7mm 0mm 2.2mm}, clip, width=0.5\textwidth]{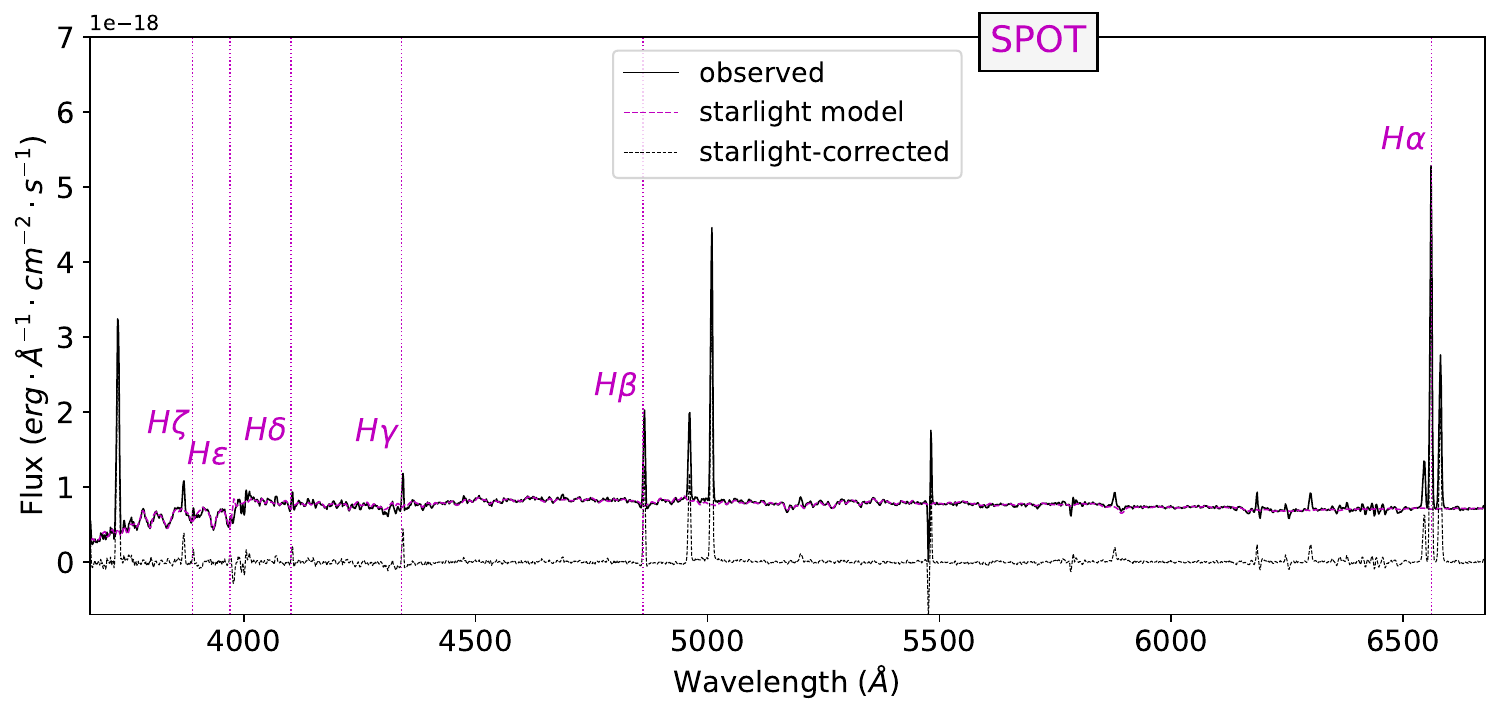}
   \caption{Direct spectra of the nucleus of GSN~069 (1.25\arcsec extraction, upper panel) and the southern spot (2.25\arcsec extraction, lower panel). The starlight model (magenta) has been subtracted from the observed spectrum (black) to generate the emission-line spectrum (dashed line).}
        \label{fig:direct_spectra}%
\end{figure}

We computed the direct spectrum of GSN~069 by adding up the 24 individual spectra taken in PMOS mode. After flux-calibrating the spectrum, by using the response curve generated with the spectropolarimetric observations of the spectrophotometric standard star Feige 110, we corrected for the galactic absorption along the line of sight throughout Cardelli's extinction law \citep{Cardelli1989}, considering $E(B-V)=0.0232\pm0.0009$, as reported by \cite{SchlaflyFinkbeiner2011} and using the python package \texttt{dust\_extinction v1.2}.\footnote{https://dust-extinction.readthedocs.io/en/stable/} Since the seeing value exceeded temporarily the slit width, we may expect a wavelength-dependent slit loss. Assuming a point-like source observed through a 1" slit with a seeing of 1.3", the considered line flux ratios will not be affected by more than 3\%, and slit loss can be neglected. 

We   estimated the host galaxy contribution to the observed spectra by modeling the stellar continuum and absorption features (e.g., MgI $\lambda\lambda$5167,5173,5184) following the same procedure as \cite{Cazzoli2018}. Briefly, we used the penalized PiXel-Fitting code (pPXF) by \citet{Cappellari2017} and references therein, with the Indo-U.S. stellar library \citep{Valdes2004}. In this library, there are 885 stars, with a continuous spectral coverage from 3460 to 9464 \AA, selected to provide us with a broad coverage of the stellar atmospheric parameters (effective temperature, surface gravity, and metallicity). We modeled the stellar component for the entire wavelength range of our spectra to avoid introducing inconsistencies or biases that affect the emission line ratios, so that the best-fit stellar spectrum can be subtracted from the observed one to obtain a pure spectrum for the nuclear emission. The nuclear spectrum (Fig.\ref{fig:direct_spectra}) shows Balmer lines in absorption from H$\beta$ onward. After starlight correction, H$\gamma$ and H$\epsilon$ clearly arise in emission along with a weak H$\delta$. We have performed line fitting around $H\alpha$ and $H\beta$ on the starlight-corrected spectra (Fig.\ref{fig:linesGSN069}). From the same line-fitting (see appendix \ref{append:broad_lines} for details), we have derived the standard emission line ratios used in the BPT diagrams (\cite{Kewley2001} and references therein). [O\,III]$\lambda$5007/H$\beta$, [N\,II]$\lambda$6583/H$\alpha$, and [O\,I]$\lambda$6300/H$\alpha$ are for the nucleus (southern spot), in logarithm: $0.94\pm0.01$ $(0.52\pm0.01)$, $0.145\pm0.005$ $(-0.27\pm0.01)$ and  $-0.77\pm0.02$ $(-1.20\pm0.06)$, which places both regions in the AGN section of the diagnostic diagrams. The direct spectra corresponding to the north and south extractions of the host galaxy are shown in Fig. \ref{fig:HOST_spectra}. BPT narrow-line ratios also indicate an AGN classification for both regions (Appendix~\ref{append:host}).

\section{Southern spot and MUSE data}
\label{append:Spot}

\begin{figure}
    \centering
   \includegraphics[scale=0.6]{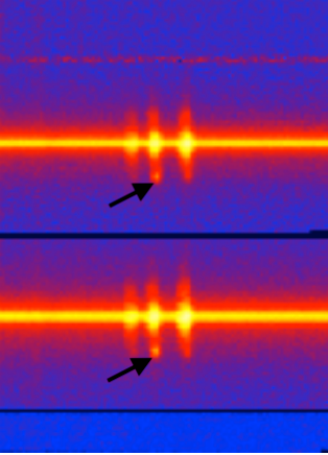}
   \caption{Bright feature located $4.25\arcsec$ south of the nucleus, which we refer to as the southern spot, as seen in the spectral images by zooming on the H$\alpha$ + [NII] spectral region. Ordinary (upper) and extraordinary (lower) spectra at one position of the HWP. The height of each strip is about 20".}
        \label{fig:spot}%
\end{figure}

In every spectral image, we detected a bright feature located $4.25\arcsec$ south of the nucleus, which we refer to as the southern spot, visible in both orthogonal polarizations and most prominent in the H$\alpha$ emission (Fig.~\ref{fig:spot}). We extracted the direct spectrum and the normalized Stokes parameters of the spot following the same procedure as for the nucleus, but using a 9-pixel (2.25$"$) subslit centered on the spot (Fig.~\ref{fig:integration_areas_contours}). The broader extraction, compared to that of the other regions, was adopted to encompass the entire feature along the spatial direction.

\begin{figure}
\centering
   \includegraphics[width=0.5\textwidth]{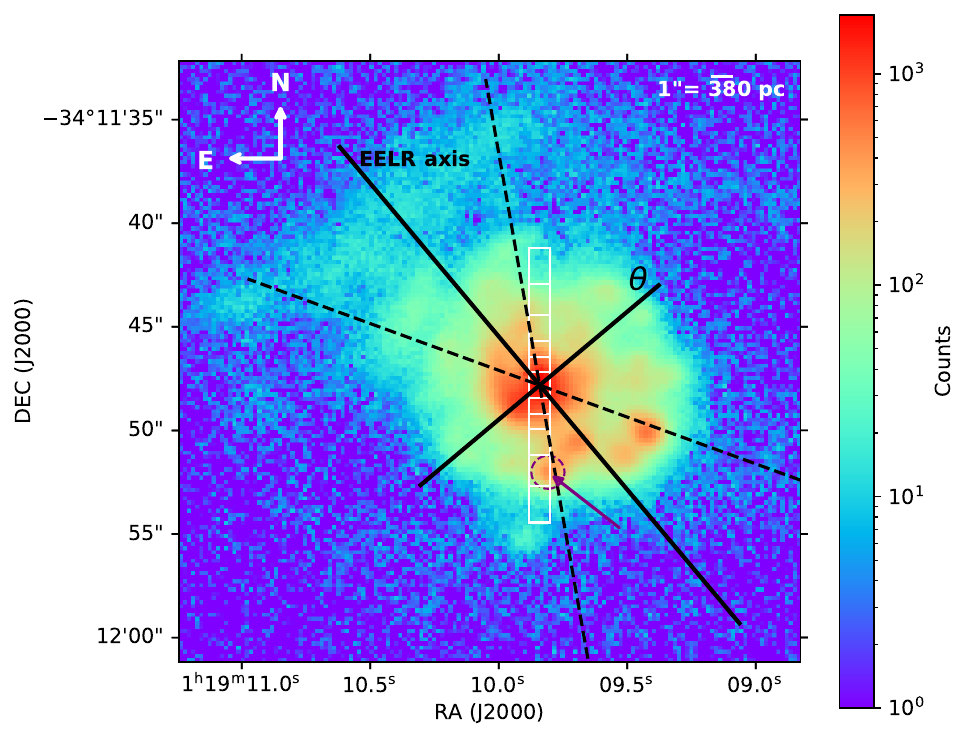} 
   \includegraphics[width=0.5\textwidth]{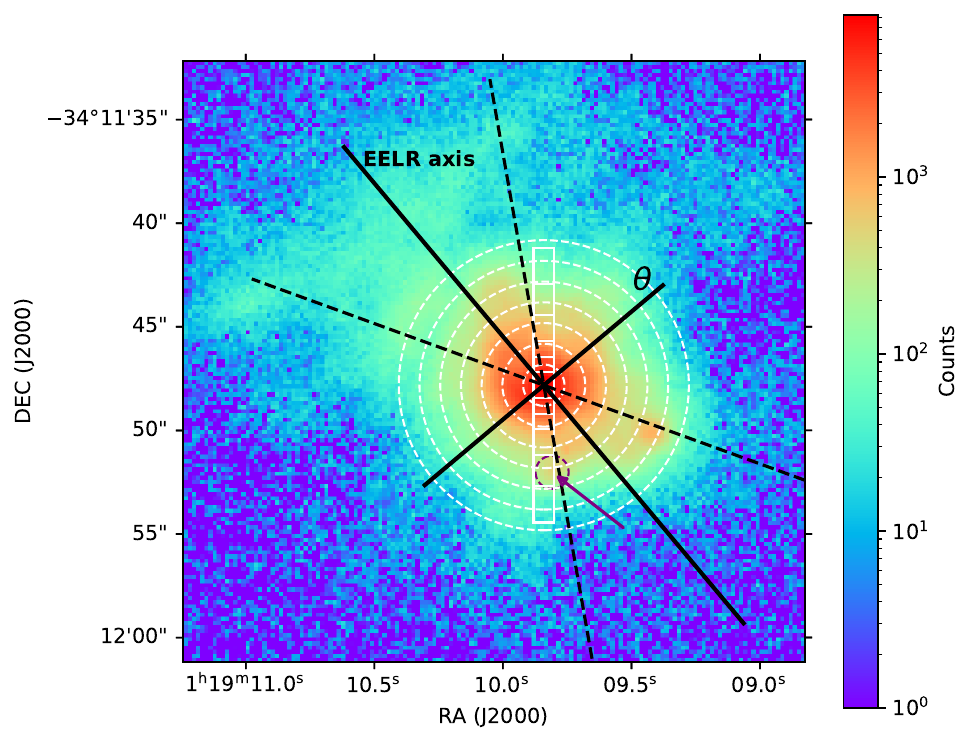}
   \includegraphics[width=0.5\textwidth]{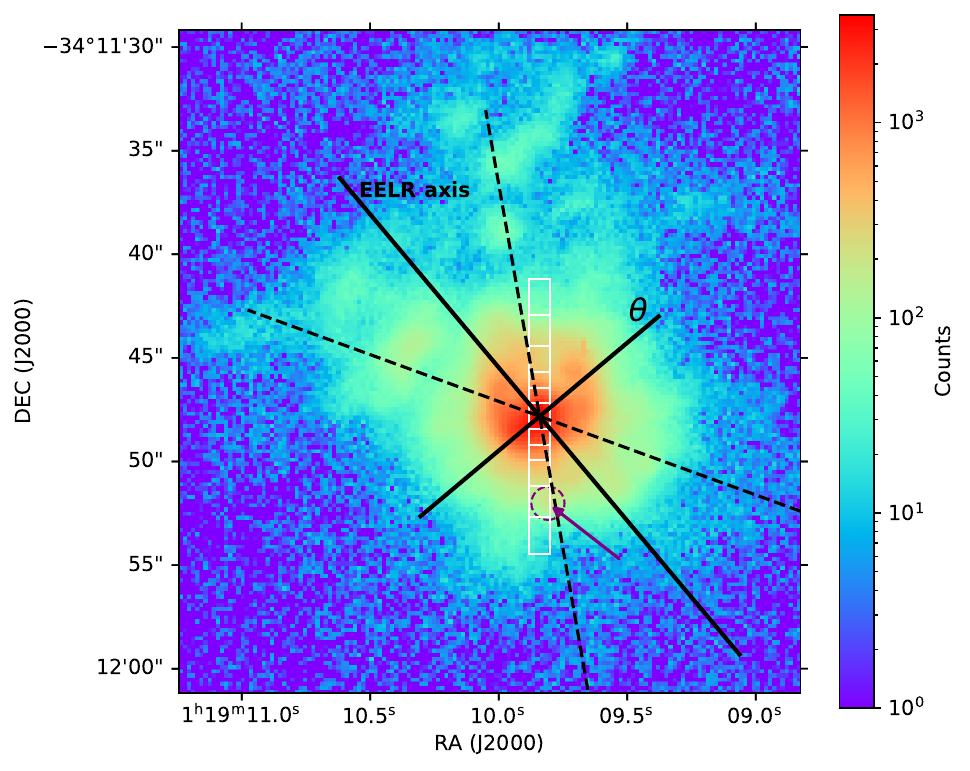}
   
   \caption{\textit{Upper panel:} MUSE continuum-subtracted H$\alpha$ image. \textit{Middle panel:} MUSE continuum-subtracted [OIII]$\lambda5007$.\textit{Lower panel:} MUSE continuum-subtracted [NII]$\lambda6583$. The position of the southern spot is marked with a purple arrow and circle. The white solid boxes denote the spectropolarimetry bins corresponding to Fig.\ref{fig:asymmetry1}, while the dashed white circles account for the annular regions of Fig.\ref{fig:asymmetry2} (only shown in the middle panel for clarity). The direction corresponding to the polarization angle $\theta$ and its perpendicular are superimposed. The latter direction represents the proposed EELR axis. The black dashed lines represent an arbitrary 60\degr aperture for the ionization cones, drawn to guide the eye.}
        \label{fig:muse}%
\end{figure}

 The spot is not seen in our optical images, but in Chandra data \citep{Miniutti2019Natur}. To further investigate its presence, we inspected the ESO archive and found the run/programme ID 109.238W.007/0109.B-0871(G) "A filler program for the apocalypse: end all weather-idle-time on UT4 in P109", PI: Bian. This program includes MUSE data of GSN~069 taken in July 2022, $27^{th}$ with a total exposure time of 3000~s and an effective spatial resolution of 1.22". Because the southern spot appears brigther in H$\alpha$ in our FORS2 spectra, we isolated that line in the MUSE data using \texttt{QFitsView 4.1}.\footnote{https://www.mpe.mpg.de/~ott/dpuser/qfitsview.html} The resulting H$\alpha$ emission map shown in Fig. \ref{fig:muse} clearly reveals the spot. Trying to resolve the EELR structure, we   also isolated the [OIII]$\lambda5007$ and [NII]$\lambda6583$ line maps (Fig. \ref{fig:muse}). In the three cases, the axis perpendicular to the polarization angle $\theta$, aligns with elongated gas distributions at the NE. This also traces a preferred direction in the illuminating photon field, favoring a geometry more consistent with a long-lived torus structure of a faded AGN than with a morphology shaped by a past intense TDE activity that is not expected to sustain nor maintain collimating structures.

\section{Host extractions north and south of the nucleus}
\label{append:host}

\begin{figure}
   \includegraphics[width=0.5\textwidth]{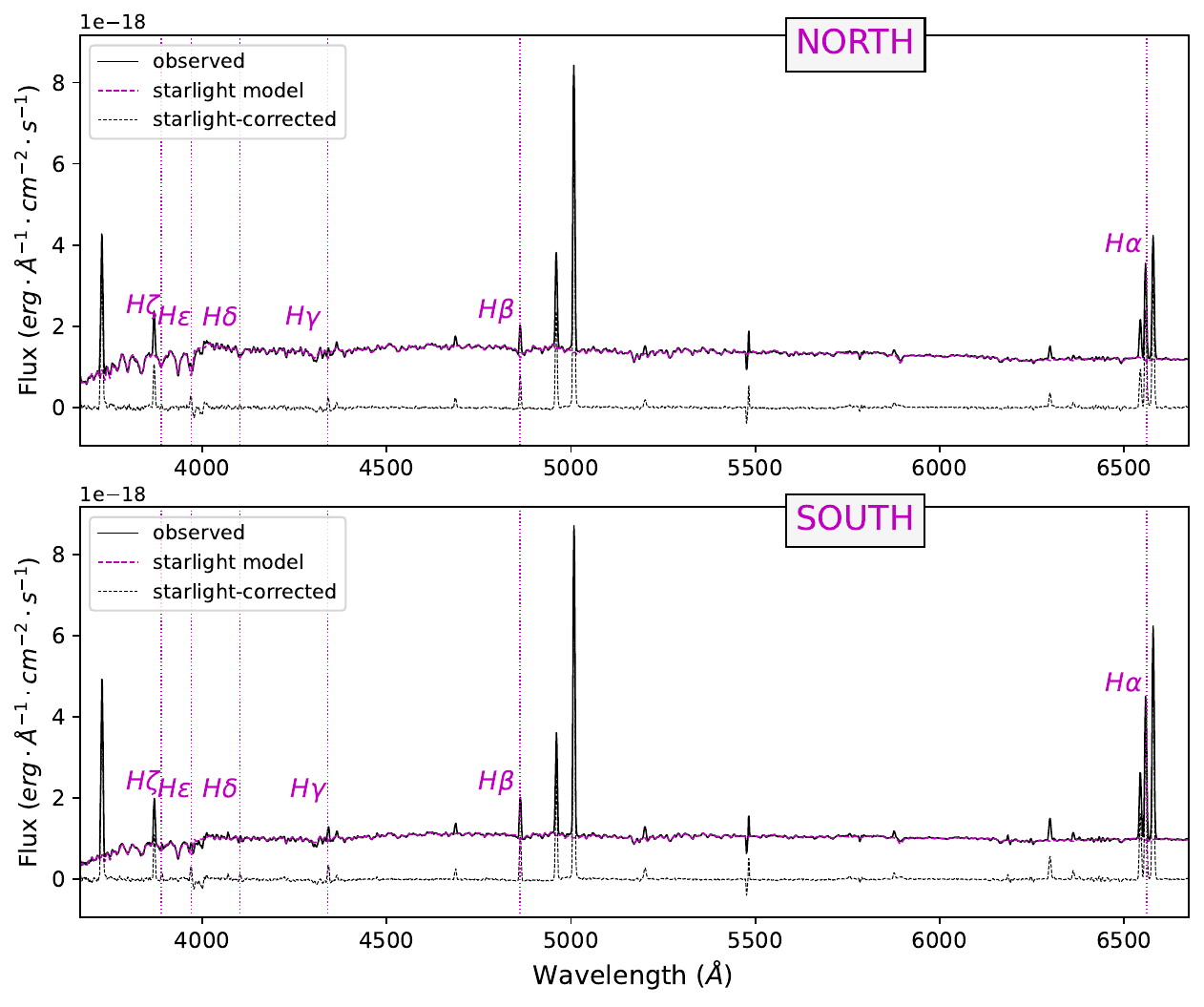} 
   \caption{Direct spectra of the regions north and south of the nucleus of GSN~069 (both 1.25\arcsec extractions). The starlight model is represented in magenta. The starlight model has been subtracted from the observed spectrum to generate the starlight-corrected spectrum shown at the bottom of each panel with a dashed line.}
        \label{fig:HOST_spectra}%
\end{figure}

The direct and polarized spectra of the host galaxy were also extracted using 5-pixel (1.25") subslits located north and south of the nucleus (Fig.~\ref{fig:integration_areas_contours}). Figure \ref{fig:HOST_spectra} shows the direct spectra for the north and south regions, with starlight subtraction done as explained in Sect.~\ref{appendix_subsec:impol_obs_red}. The line ratios in the north (south) region are, in logarithm: [OIII]$\lambda$5007/H$_{\alpha}= 0.95\pm0.02\ (0.85\pm0.02)$, [N\,II]$\lambda6583/H_{\alpha}= 0.11\pm0.01\ (0.157\pm0.005)$, and [O\,I]$\lambda$6300/H$_{\alpha}=0.85\pm0.03\ (-0.80\pm0.03)$, which in the BPT diagrams translates into AGN classifications for both regions. 

\begin{figure}
   \centering 
   \includegraphics[scale=0.30]{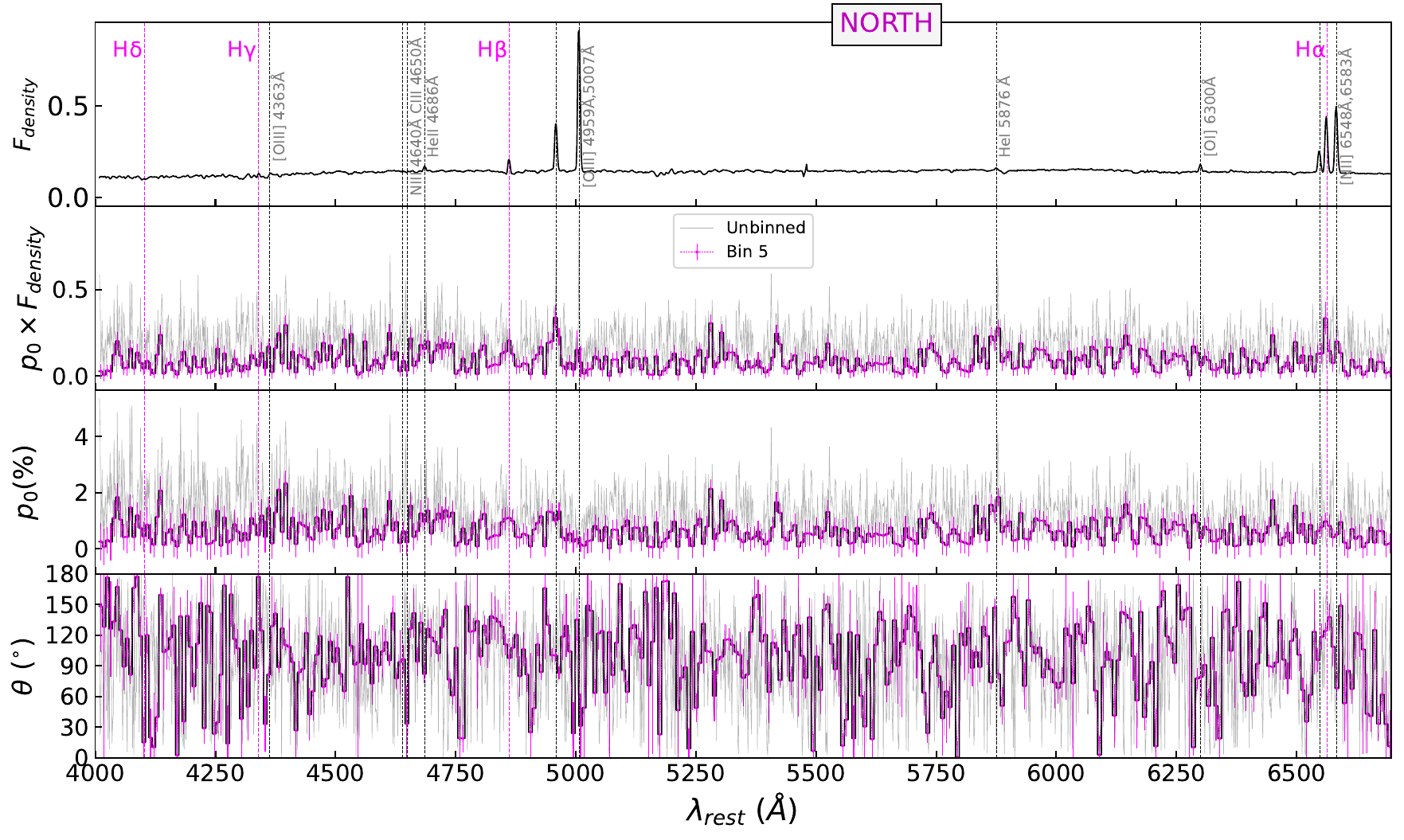}
   \includegraphics[scale=0.30]{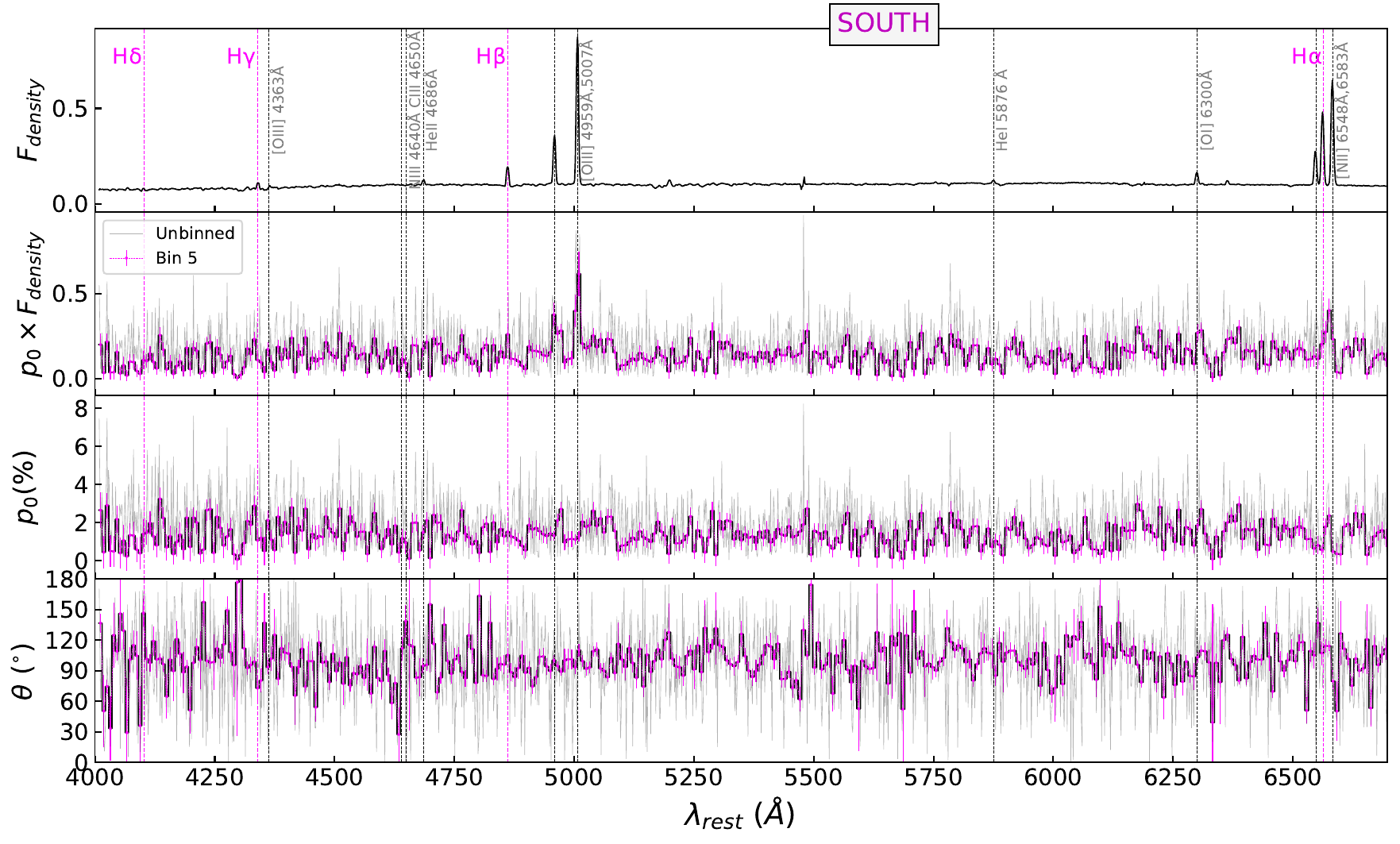}
   \caption{Flux density in arbitrary units ($F_{density}$), polarized flux ($F_{density}\times p_{0}$), debiased polarization degree ($p_{0}$), and polarization angle ($\theta$) for the north (upper panel) and south (lower panel) regions.}
   \label{fig:specpolGSN069host}%
\end{figure}

Figure \ref{fig:specpolGSN069host} shows the polarized spectra for the north and south regions.

\section{Broad line fitting}
\label{append:broad_lines}
\begin{figure}
\centering 
\includegraphics[scale=0.3]{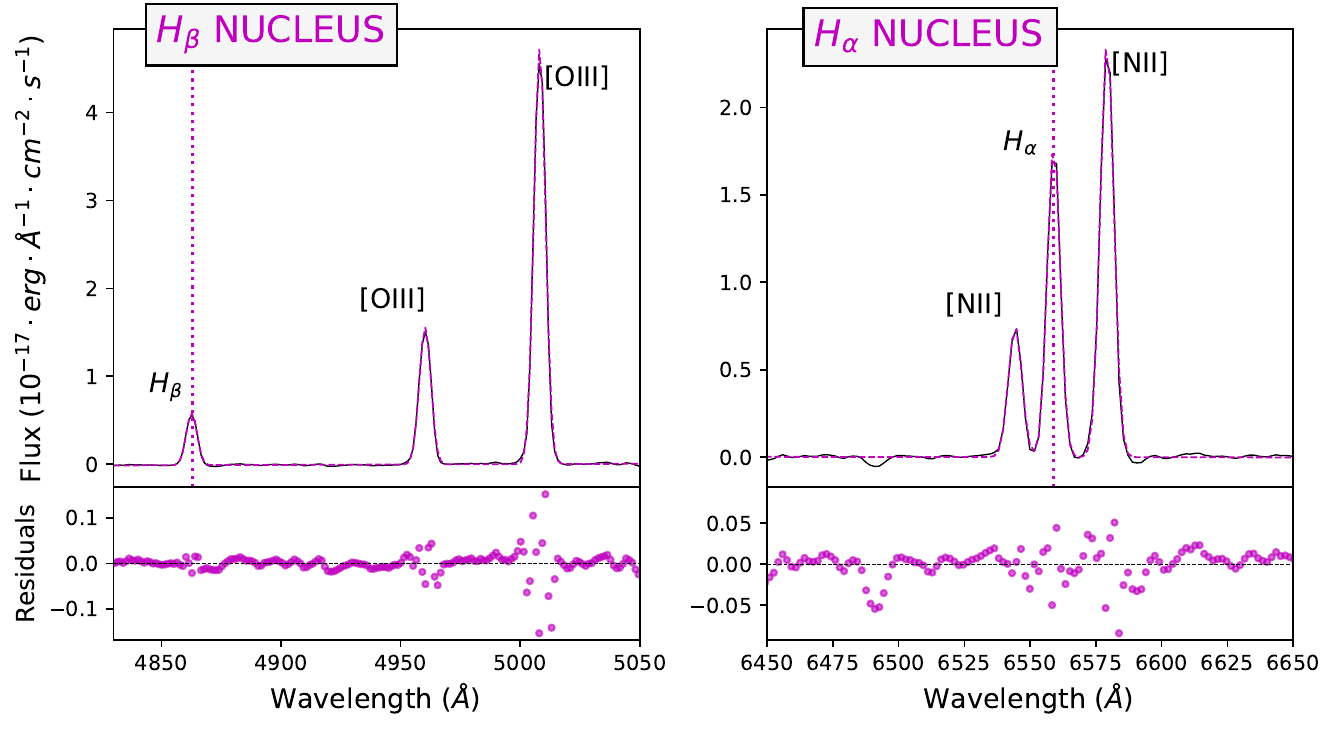}
\includegraphics[scale=0.3]{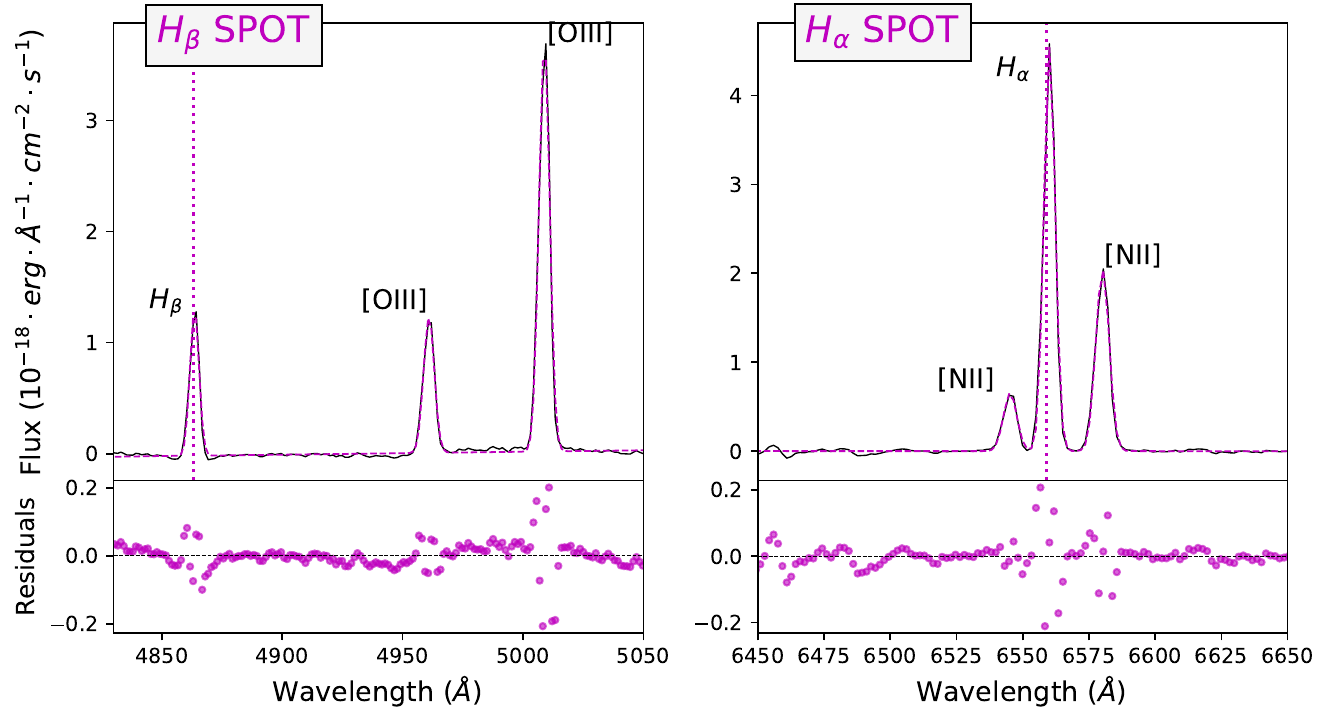} 

   \caption{Line fitting and corresponding residuals  in the H$\beta$ and H$\alpha$ wavelength ranges for the direct nucleus spectrum of GSN~069 (upper panel) and the southern spot (lower panel)).}
        \label{fig:linesGSN069}%
\end{figure}

Figure \ref{fig:linesGSN069} shows the line fitting around H$\alpha$ and H$\beta$ for the direct spectra of nucleus of GSN069 and the southern spot. Narrow lines were modeled as single Gaussians (Lorentzians provided worst residuals). Spectral lines belonging to the same doublet were assumed to have the same FWHM. A second broad Gaussian was added to check if the statistic improves, using \texttt{curve\_fit}, but that is not the case, while the results are not physically meaningful. As common in optical emission-line fitting, underestimated pixel uncertainties give meaningless reduced $\chi^2$, in this case with and without broad Gaussian $\sim13$. The covariance matrix has been re-scaled by the reduced $\chi^2$ factor to obtain realistic parameter uncertainties.

\begin{figure}
   \centering 
   \includegraphics[scale=0.28]{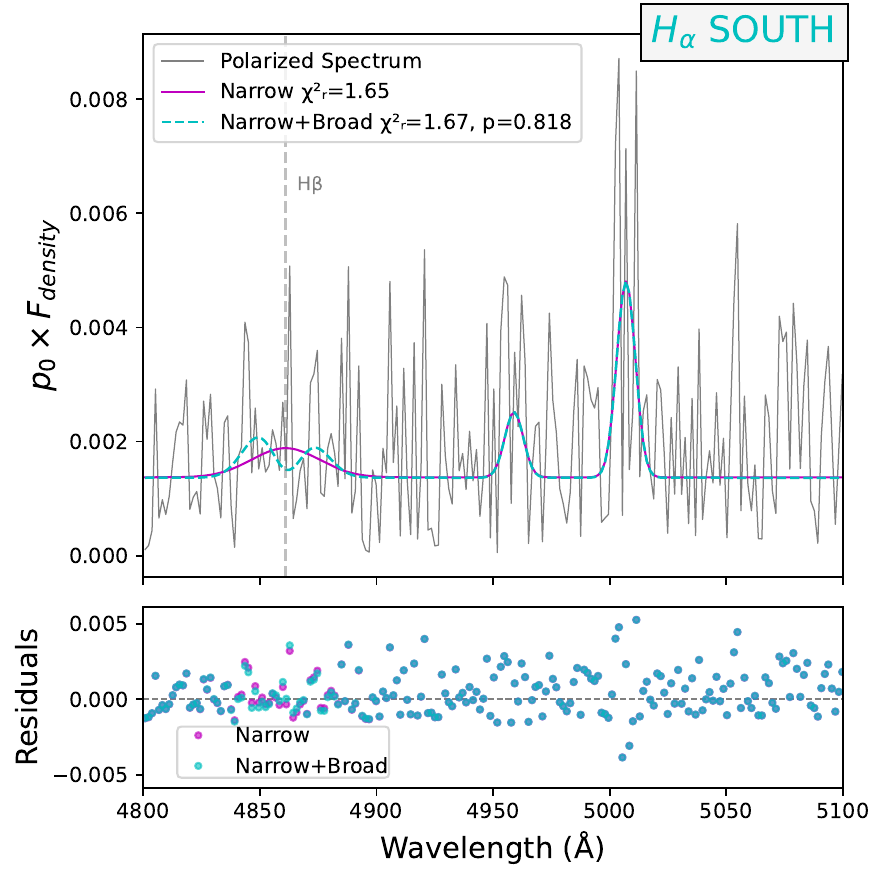}
   \includegraphics[scale=0.28]{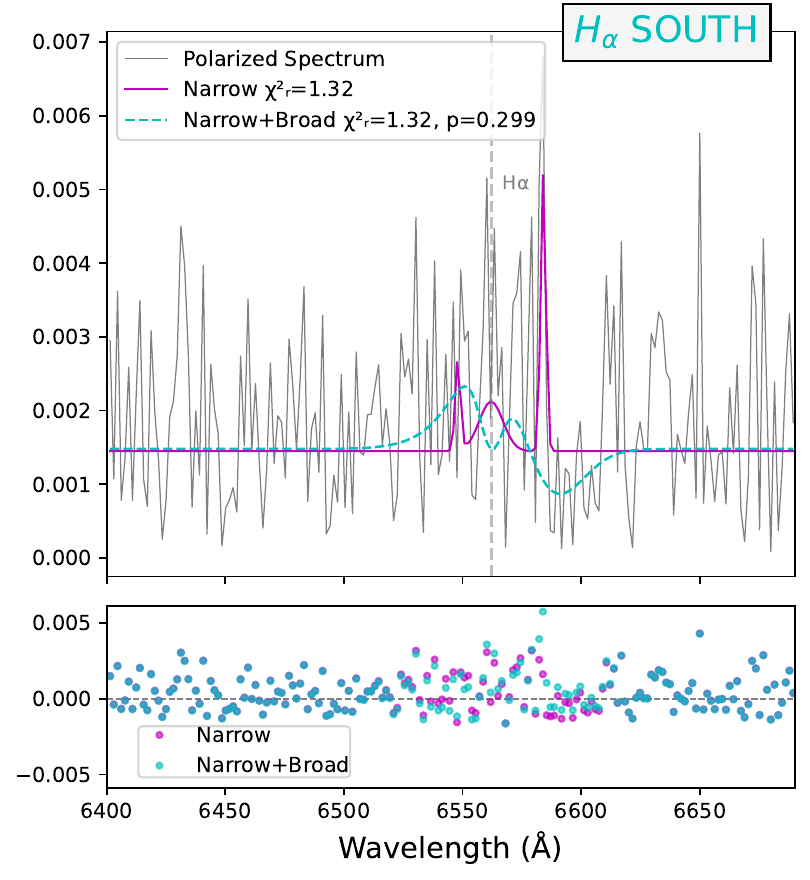}
   \includegraphics[scale=0.28]{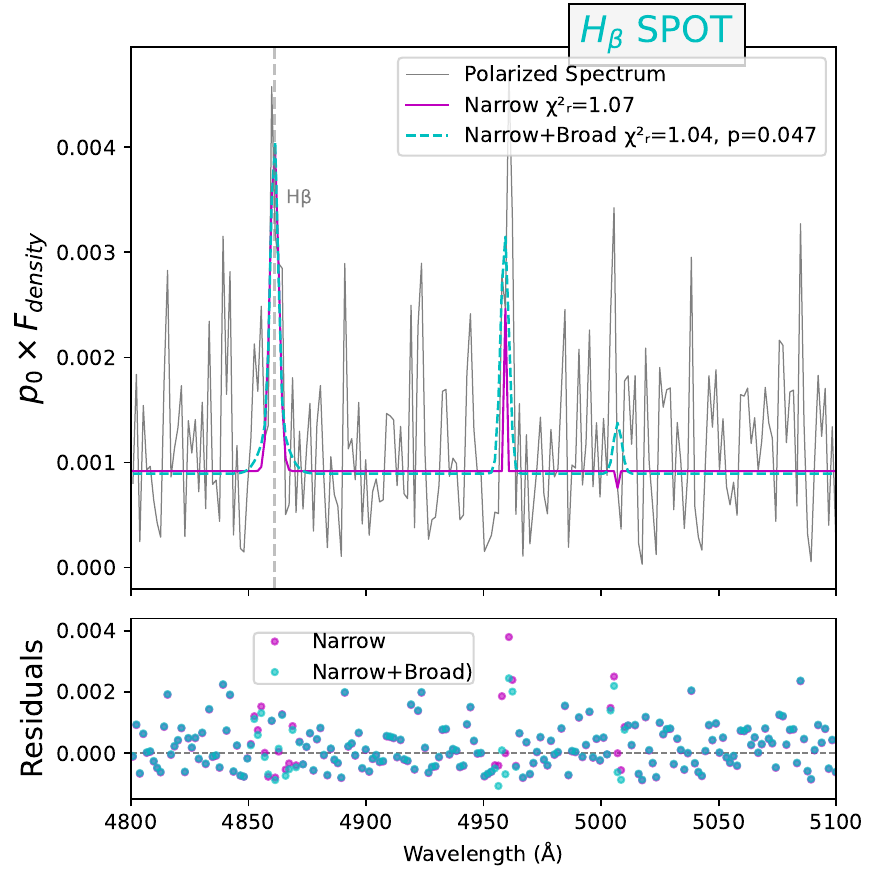}
   \includegraphics[scale=0.28]{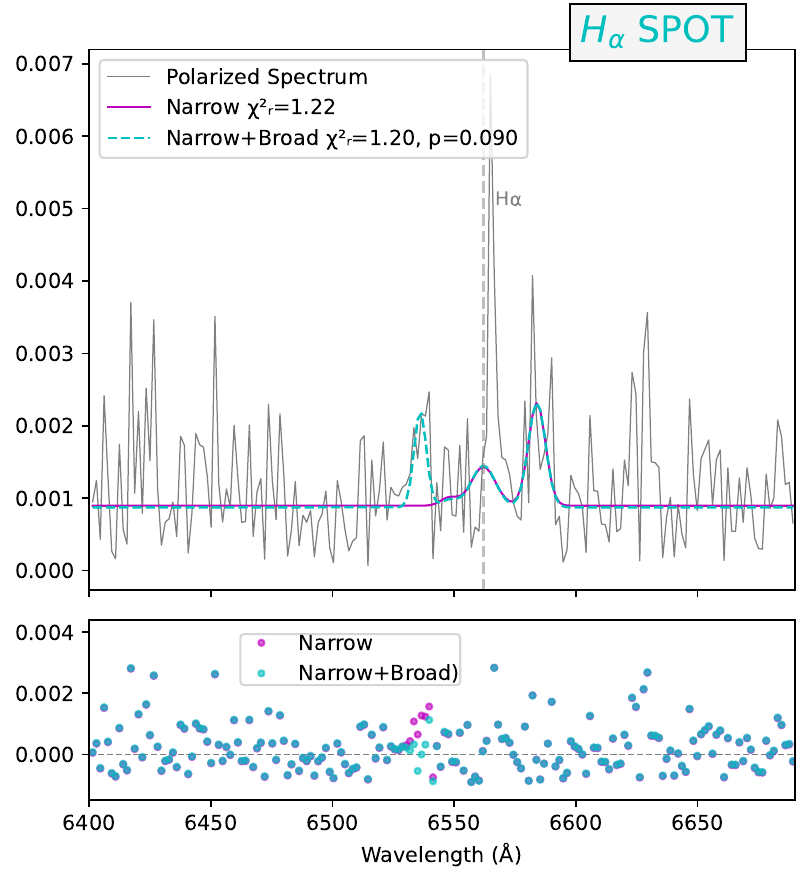}
   \caption{Line fitting and corresponding residuals for the polarized spectrum of the south (upper row) and the southern spot (lower row) extractions in the H$\beta$ and H$\alpha$ wavelength ranges. As shown in the inset, the addition of broad emission line components is never significant.}
   \label{fig:fits_polarized_spectrum}%
\end{figure}

Since the south extractions are the ones with higher polarization, we also carried out line fitting on the corresponding polarized spectra, even if those are noisy, to rule out the presence broad Balmer components there. With the polarized spectra, if broad Gaussians were tied to the narrow component, the fitting did not converge. Nonetheless, offsets between BELs and NELs are seen in AGNs \citep{Gaskell1982}. In the south extraction, the reduced $\chi^{2}$ does not improve by adding broad Gaussians, which are neither physically meaningful (see figure \ref{fig:fits_polarized_spectrum} for the statistics). Applying the same fitting for the southern spot (figure \ref{fig:fits_polarized_spectrum}) the broad components are neither statistically required nor physically justified.

\end{appendix}

\begin{acknowledgements}

BAG, IL were funded by the European Union ERC-2022-STG - BOOTES - 101076343. Views and opinions expressed are however those of the author(s) only and do not necessarily reflect those of the European Union or the European Research Council Executive Agency. Neither the European Union nor the granting authority can be held responsible for them. BAG, SC, IM, JM acknowledge financial support from the research project PID2022-140871NB-C21 from MICIU/AEI/10.13039/501100011033 and FEDER, UE, through the Center of Excellence Severo Ochoa award for the Instituto de Astrof\'isica de Andaluc\'ia-CSIC (CEX2021-001131-S). BAG acknowledges support provided by the Fonds de la Recherche Scientifique - FNRS, Belgium, under grant No. 4.4501.19 during the fisrt stage of this work. DH is Research Director F.R.S.-FNRS. GM acknowledges support from grants n. PID2020-115325GB-C31 and n. PID2023-147338NB-C21 funded by MCIN/AEI/10.13039/50110001103. CRA, JAP acknowledge support from the Agencia Estatal de Investigación of the Ministerio de Ciencia, Innovación y Universidades (MCIU/AEI) under the grant “Tracking active galactic nuclei feedback from parsec to kiloparsec scales”, with reference PID2022-141105NB-I00 and the European Regional Devel- opment Fund (ERDF). Based on observations collected at the European Organisation for Astronomical Research in the Southern Hemisphere under ESO programme 0103.B-0791. We thank the referee for their enriching comments and discussion that have improved the manuscript quality.

\end{acknowledgements}

\end{document}